\documentclass[twocolumn,prb,showpacs]{revtex4}
\usepackage{graphicx}
\usepackage{dcolumn}
\newcommand{\bm}[1]{\mbox{\boldmath$#1$}}
\begin{document}
%
\title{
 Quarter-filled extended Hubbard model with alternating transfer
 integral:\\ Two-dimensional Ising transition in the ground state 
}
\author{
 Hiromi Otsuka$^{1}$ and Masaaki Nakamura$^{2}$
}
\address{
 $^1$Department of Physics, Tokyo Metropolitan University, Tokyo
 192-0397 Japan\\
 $^2$Department of Applied Physics, Faculty of Science, Tokyo University
 of Science, Tokyo 162-8601 Japan
} 
\date{\today}
\begin{abstract}
 We study the one-dimensional quarter-filled extended Hubbard model with
 an alternating transfer integral.
 In the strong-dimerization limit the charge part is described by the
 quantum Ising model which shows the two-dimensional Ising criticality
 at the self-dual point, and it is naturally connected to the
 double-frequency sine-Gordon theory in the weak dimerization.
 Treating low-lying excitations in finite-size systems, we numerically
 determine a phase boundary between two types of $4k_{\rm F}$
 density-wave states and clarify the ground-state phase diagram.
 Further, we refer to its relevances to the charge-ordered phase
 observed in the charge-transfer organic salts.
\end{abstract}
\maketitle


 The organic materials described by the chemical formula (TMTSF)$_2$X
 (X=PF$_6$, ClO$_4$, etc) and (TMTTF)$_2$X (X=PF$_6$, AsF$_6$, etc)
 form a class of the quasi one-dimensional (1D) conductors; a large
 number of investigations on these materials have been accumulated in 
 the literature.\cite{Bour99}
 While the various types of electronic phases, e.g.,
 spin/charge-density-wave (SDW/CDW), the spin-Peierls and the
 superconducting states have been observed in the low-temperature
 region, newly discovered charge-ordered (CO) phase in (TMTTF)$_2$X
 exhibiting an anomaly in the low-frequency dielectric constant
 \cite{Nad_00}
 and the charge disproportionation in the NMR studies \cite{Chow00}
 has received intensive current interest.
 Although the stabilizations of these phases at finite temperature
 resort to interchain couplings, it is believed that the intrachain
 interaction effects play a leading role to describe them.

 For the study of the CO phase, the 1D quarter-filled extended Hubbard
 model (EHM) with an alternating transfer integral has been
 employed:\cite{Mila95,Seo_97,Mazu99,Tsuc01,Tsuc02,Nish00,Shib01}
 $H=H_1+H_2$ with 
 \begin{eqnarray}
  &&H_1=
   \sum_{j,s}
   -t\left[1-\delta(-1)^j\right]
   \left(
    c^\dagger_{j,s}c^{}_{j+1,s}+{\rm H.c.}
   \right),\\
  &&H_2=
   \sum_j \left(U n_{j,\uparrow} n_{j,\downarrow}+ V n_{j} n_{j+1}\right),
 \end{eqnarray}
 where $c^{}_{j,s}$ annihilates an $s$-spin electron ($s=\uparrow$ or
 $\downarrow$) on the $j$th site and satisfies the periodic boundary
 condition $c^{}_{L+1,s}=c^{}_{1,s}$ ($j\in[1,L]$; $L$ is an even
 number).
 The number operators are defined as 
 $n_{j,s}=c^\dagger_{j,s}c^{}_{j,s}$ and
 $n_{j}=n_{j,\uparrow}+n_{j,\downarrow}$. 
 The parameters $U$ and $V$ taking positive values stand for the onsite
 and nearest-neighbor Coulomb repulsion. 
 The dimerization parameter $\delta$ shows the alternation in the transfer
 integral of the molecular chains (we set $t=1$ in the following).
 For the theoretical descriptions of the 1D electrons, the
 Tomonaga-Luttinger liquid (TLL) picture has been widely
 adopted.\cite{Hald81} 
 Since TLL consists of the massless charge and spin parts both
 controlled by the Gaussian fixed point [the conformal field theory
 (CFT) with the central charge $c=1$], it is important to understand its
 instabilities.
 In particular, the CO transition may be related to the crossover of the
 criticality embedded in the renormalization group (RG)
 flow,\cite{Tsuc02}
 which is one of the typical instability of the $c=$1 CFT.

 In this paper, we present the numerical calculation results of the
 ground-state phase diagrams of $H$.
 Our method being deeply connected to the instability will be explained
 briefly.
 Further, in the strong-dimerization limit ($\delta=1$), we show that
 the charge part can be described by the so-called quantum Ising chain,
 which is complementary to the bosonization argument and gives us an
 exact limiting condition of the phase boundary line.
 Since the region with sufficiently large Coulomb repulsions is relevant
 to the CO transition, an occurrence of the phase separation
 or the transition to the superconducting phases is outside of our 
 research scope.

 Let us start with the description of the low-energy physics in the
 weak-coupling region, where the bosonization method provides a reliable
 approach, 
 i.e., linearizing the dispersion at Fermi points $\pm k_{\rm F}=\pm\pi
 n/2a$ (electron density $n=N/L=1/2$) and applying the method, we can
 obtain an effective Hamiltonian.
 For the present case, according to the recent research
 results,\cite{Bour99,Tsuc01,Yosh00} 
 we can use the following expression:
 $H\rightarrow{\cal H}={\cal H}_\rho+{\cal H}_\sigma$
 with 
 \begin{eqnarray}
  {\cal H}_\rho
   \!\!&=&\!\!
   \int {d}x \frac{v_\rho}{2\pi}
   \left[
    {       K_\rho}\left(\partial_x \theta_\rho\right)^2+
    {1\over K_\rho}\left(\partial_x   \phi_\rho\right)^2 
  \right]\nonumber\\
  \!\!&+&\!\!
   \int {d}x \frac{2}{(2\pi\alpha)^2}\!
   \left(
    -g_\rho{\sin\sqrt8\phi_\rho}+g_{1/4}{\cos2\sqrt{8}\phi_\rho}
  \right),
   \label{eq_chag}
 \end{eqnarray}
 where the operator $\theta_\rho$ is the dual field of $\phi_\rho$
 satisfying the commutation relation
 $\left[\phi_\rho(x),\partial_y\theta_{\rho}(y)/\pi\right]={i}\delta(x-y)$
 and parameters $K_\rho$ and $v_\rho$ are the Gaussian coupling and the
 velocity of the charge excitation, respectively.\cite{GNT}
 A benefit to use the bosonized expression is now clear, i.e., since the
 spin-charge separation occurs in ${\cal H}$ and ${\cal H}_\sigma$ is
 the SU(2) critical Gaussian model in the present
 case,\cite{Tsuc01,Tsuc02} we can concentrate on the charge part ${\cal
 H}_\rho$ which takes a form of the so-called double-frequency
 sine-Gordon (DSG) model.
 In uniform case ($\delta=0$), the $8k_{\rm F}$-Umklapp scattering with
 $g_{1/4}\propto U^2(U-4V)$ (Refs.\
 \onlinecite{Yosh00,Yosh01,Giam97,NakaEX}) brings about the
 Berezinskii-Kosterlitz-Thouless (BKT) transition, and then the charge
 part becomes massive for large values of the Coulomb interactions.
 For the BKT transition point, values in the strong coupling limit are
 known as $V^*(U\to\infty)=2$ and
 $U^*(V\to\infty)=4$.\cite{Ovch73,Lin-G-C-F-G}
 Further the estimations for the intermediate region are
 available.\cite{Yosh00,Mila93,NakaEX}
 In the case of nonzero dimerization ($\delta\ne0$), the scaling
 dimension of the ``half-filled Umklapp scattering'' term with
 $g_{\rho}\propto U\delta[1-A(U-2V)]$ ($A$ is a constant)
 \cite{Tsuc01,Giam97} on the Gaussian fixed point is small ($x_{\rm
 4B}=2K_\rho$) enough to bring about the second-order phase transition
 for $V\le V^*(U)$, which is accompanied by the divergent correlation
 length of the form $\xi\propto
 \delta^{-1/(2-2K_\rho)}$.\cite{Bour99,Penc94}
 For $V>V^*(U)$, since the charge gap may survive in a weak-dimerization
 region, the transition point $\delta_\rho(U,V)$ takes nonzero values
 depending on $U$ and $V$, and more importantly, the universality of the
 transition is changed.
 Recently, Tsuchiizu and Orignac,\cite{Tsuc02}
 on the basis of the DSG theory,\cite{Delf98}
 argued that the charge part on $\delta_\rho(U,V)$ [$V>V^*(U)$] is
 renormalized to the 2D-Ising fixed point with $c=\frac12$ (i.e., the
 fixed point with lower symmetry), which is in accord with
 Zamolodchikov's $c$-theorem \cite{Zamo86}
 (see also Refs.\ \onlinecite{Fabr99} and \onlinecite{Fabr00}).
 Then, the critical line corresponds to the phase boundary and satisfies
 a condition $\delta_\rho(U,V\to V^*(U))\searrow 0$ in the
 weak-dimerization region.
 To characterize the phases, we shall use the CDW and the
 bond-order-wave (BOW) order parameters with the $4k_{\rm F}$ wave
 vector:\cite{Yosh00} 
 \begin{equation}
  {\cal O}_{\rm 4C} \propto \cos\sqrt8\phi_\rho,~~~
  {\cal O}_{\rm 4B} \propto \sin\sqrt8\phi_\rho. 
 \end{equation}
 Here, note that the expectation value of the $4k_{\rm F}$-BOW order
 parameter is finite,
 $\langle{\cal O}_{\rm 4B}\rangle\ne0$ and
 $\langle{\cal O}_{\rm 4C}\rangle  =0$ in the upper region of the
 boundary, but both of these are finite in the lower region ($\delta\ne0$).
 While this ``mixed'' state is basically the $4k_{\rm F}$-CDW phase, we
 shall use the double quotation marks ``$4k_{\rm F}$-CDW'' to express
 this situation.\cite{Fabr00}

 On the other hand, another condition of the boundary can be found in
 the strong-dimerization limit ($\delta=1$).
 To derive an effective Hamiltonian, it is convenient to work with the
 orbital operators defined by $ d^{}_{m,\pm,s}
 \equiv \left(c^{}_{2m-1,s}\pm c^{}_{2m,s}\right)/{\sqrt2} $, where
 $d^{}_{m,l,s}$ annihilates an $s$-spin electron in the $l$-orbital
 ($l=\pm$) on the $m$th unit cell ($m\in[1,L/2]$).
 In this limit, $H_1$ consists of a sum of the intracell electron
 hopping, which is diagonalized by using the operators as
 $H_1=\sum_{m,l,s}-2ld^{\dagger}_{m,l,s}d^{}_{m,l,s}$.
 For sufficiently large $U$ and $V$, since the one-electron states
 $|l,s\rangle_m=d^\dagger_{m,l,s}|0\rangle$
 have a principal role to describe the $m$th unit cell in the
 quarter-filled ground state, and the Hamiltonian does not change the
 electron number in each cell, we shall introduce the pseudospin
 operators,
  \begin{equation}
  {\bm T}_m
   \equiv
   \sum_{l,l',s}
   \frac12
   d^\dagger_{m,l,s }
   \left[\mbox{\boldmath$\tau$}\right]^{}_{l,l'}
   d^{     }_{m,l',s}, 
  \end{equation}
 acting on the orbital space as, for instance, $T^3_m |\pm,s\rangle_m =
 \pm\frac12 |\pm,s\rangle_m$ 
 [$\mbox{\boldmath$\tau$}=(\tau^1,\tau^2,\tau^3)$; $\tau^i$ is the Pauli
 matrix].
 Using these, $H_1=\sum_{m}-4T^3_m$.
 For $H_2$, since the intracell Coulomb interactions are absent and the
 intercell Coulomb repulsion only remains in the restricted Hilbert
 space spanned by the direct product of one-particle states
 $\{\otimes_m|l,s\rangle_m\}$, a straightforward calculation brings
 about the expression
 $H_2=\sum_{m} \left(-V T^1_mT^1_{m+1}+{\rm const}\right)$.
 Now, since the Hamiltonian acts only on the orbital space, its
 eigenstate takes a form of the direct product of vectors in the spin
 and the orbital spaces as $|\Phi\rangle=|{\rm spin}\rangle\otimes|{\rm
 orbital}\rangle$.
 Thus, assuming a certain spin configuration belonging to the
 2$^{L/2}$-dimensional space for spins and restricting ourselves
 to the orbital (or charge) part, we see that the Hamiltonian $H$ with
 $\delta=1$ is reduced to the quantum Ising chain \cite{Pfeu70}
 \begin{eqnarray}
  H_{\rho,\delta=1}=\sum_m
   \left(-\Gamma T^3_m-JT^1_mT^1_{m+1}\right)
   \label{Eq_qim}
 \end{eqnarray}
 ($\Gamma=4$, $J=V$).
 Note that this possibility was mentioned qualitatively in Ref.\ 
 \onlinecite{Shib01}. 
 Then, the ground state of Eq.\ (\ref{Eq_qim}) is known to show the
 2D-Ising criticality at its self-dual point
 $\Gamma=J/2$ ($V=8)$,
 which separates ordered
 ($\langle T^1_m\rangle\ne0$)
 and disordered
 ($\langle T^1_m\rangle  =0$)
 phases.
 The ordered state is realized via the breaking of the Z$_2$ symmetry
 $(\tau^1\!\to-\tau^1)$, and it is doubly degenerated, e.g., 
 \begin{eqnarray}
  |\pm\tau^1\rangle
   =
   \prod_m\frac{1}{\sqrt2}(d^{\dag}_{m,+}\pm d^{\dag}_{m,-})|0\rangle
   =
   \prod_m c^{\dagger}_{2m-1} (c^{\dagger}_{2m})|0\rangle 
   \label{4C}
 \end{eqnarray}
 (we dropped the spin index). 
 This expresses the 4$k_{\rm F}$-CDW state with the perfect microscopic
 polarization,
 $
 \langle\pm\tau^1|T^1_m|\pm\tau^1\rangle
 =
 \langle\pm\tau^1|\frac12(n_{2m-1}-n_{2m})|\pm\tau^1\rangle
 =
 \pm\frac12
 $.
 On one hand, a disordered state is supported by the external field in
 $\tau^3$-direction, and an ideal one is given by
 \begin{eqnarray}
  |+\tau^3\rangle
   =
   \prod_m d^{\dag}_{m,+}|0\rangle
   =\prod_m\frac{1}{\sqrt2}(c^{\dagger}_{2m-1}+c^{\dagger}_{2m})|0\rangle, 
   \label{4B}
 \end{eqnarray}
 which expresses the 4$k_{\rm F}$-BOW state as expected.
 Here it is worthy of noticing that these states can be distinguished
 by the expectation value of the twist operator\cite{Naka02}
 \begin{equation}
 z_\rho\equiv\Bigl\langle{\rm exp}
  \Bigl(\frac{4\pi i}{L}\sum_j jn_j\Bigr)\Bigr\rangle.
 \end{equation}
 This quantity takes values $z_\rho=1$ for Eq.\ (\ref{4C}) and
 $z_\rho=-[\cos(2\pi/L)]^{L/2}$ for Eq.\ (\ref{4B}), so the sign of
 $z_\rho$ characterizes these two density-wave states (see below).
 Consequently, in the strong-dimerization limit, the orbital degrees of
 freedom show the 2D-Ising type transition between the ``4$k_{\rm
 F}$-CDW'' and the 4$k_{\rm F}$-BOW phases at $V=8$, where $U$ is
 irrelevant.
 Since this pseudospin representation is naturally connected to the
 bosonization picture in the weak couplings,\cite{Tsuc02}
 the phase boundary belongs to the 2D-Ising universality and satisfies
 the limiting condition $\delta_\rho(U,V\to 8)\nearrow 1$, which provides a
 solid guide to investigations in the strong-dimerization region. 


 Here, note that the qualitative estimation of the phase boundary might
 be possible in the weak- and strong-dimerization region.\cite{Tsuc01}
 To evaluate the entire phase diagram precisely, however, a numerical
 treatment of the 1D electron model is required.
 For this issue, recently the present authors have numerically treated
 the same instability observed in the quantum-spin chain and interacting
 electron systems.\cite{Otsu02}
 Therefore, we shall employ the same approach to the present system (see
 also Ref.\ \onlinecite{Kita97}).
 Since there are two critical fixed points connected by the RG flow, a
 relationship between lower-energy excitations on these fixed points
 --- the ultraviolet-infrared (UV-IR) operator correspondence ---
 has essential significance in the investigations.\cite{Tsuc02,Fabr00}
 To see this, let us rescale phase fields and the Gaussian coupling as
 $2\phi_\rho\to\phi$,
 $\theta_\rho/2\to\theta$, and 
 $4K_\rho\to K\simeq 1$, which makes it possible to directly adopt our
 previous research.\cite{Otsu02}
 With respect to $\phi$, the nonlinear potential density is given as
 $-g_\rho\sin\sqrt2\phi+g_{1/4}\cos\sqrt{8}\phi$,
 and the order parameters as
 ${\cal O}_{\rm 4C}\propto\cos\sqrt2\phi$ $(x_{\rm 4C}=K/2)$
 and
 ${\cal O}_{\rm 4B}\propto\sin\sqrt2\phi$ $(x_{\rm 4B}=K/2)$.
 Along the RG flow these operators on the Gaussian fixed point (UV) are
 transmuted to those on the 2D-Ising fixed point (IR) as
 \begin{equation}
  {\cal O}_{\rm 4C} \to \mu,~~~
  {\cal O}_{\rm 4B} \to I+\epsilon, 
 \end{equation}
 where $\mu$ is the disorder field (Z$_2$ odd), and $\epsilon$ is the
 energy density operator (Z$_2$ even) with scaling dimensions
 $x_\mu=\frac18$ and $x_\epsilon=1$, respectively.
 Since the dimerization $\delta$ couples with ${\cal O}_{\rm 4B}$ in the 
 Hamiltonian (\ref{eq_chag}), a deviation from the transition point
 $\delta-\delta_\rho(U,V)$ plays a role of the ``thermal scaling variable'' and
 brings about $\xi\propto [\delta-\delta_\rho(U,V)]^{-\nu}$ with
 $1/\nu=2-x_\epsilon=1$.
 On one hand, the operator $\mu$ corresponding to ${\cal O}_{\rm 4C}$
 provides a most divergent fluctuation.

\begin{figure}[t]
 \includegraphics[width=3.29in]{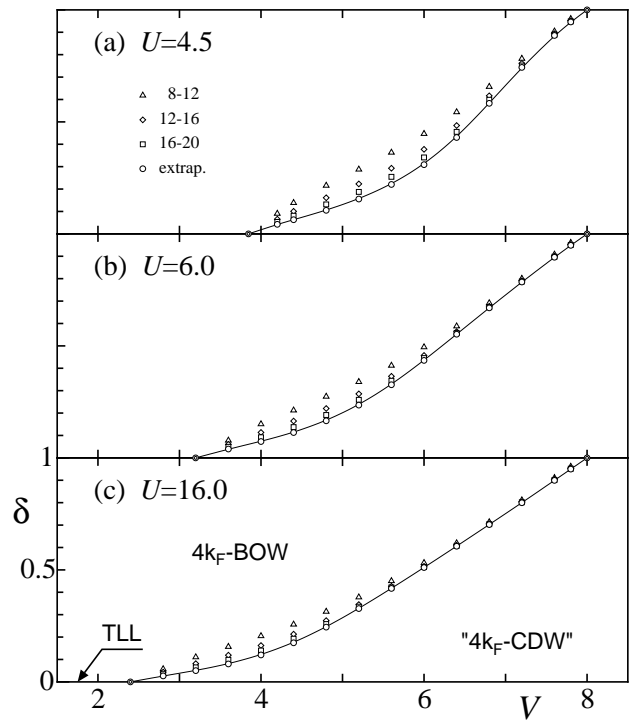}
 \caption{
 The ground-state phase diagram of the quarter-filled EHM with an
 alternating transfer integral.
 The correspondence between marks and system sizes is given in the
 figure.
 The double circles show the limiting values, i.e.,
 $(V^*(U),0)$ at which criticality changes from the Gaussian to
 the 2D-Ising type, and $(8,1)$ the self-dual point.
 }
 \label{FIG1}
\end{figure}

 Now, we shall explain our numerical procedure to determine the
 transition point.
 We shall focus our attention on the level $\Delta E$ in finite-size
 systems which corresponds to the operator ${\cal O}_{\rm 4C}$ (taking
 the ground-state energy as zero).
 According to the finite-size-scaling argument based on CFT, $\Delta
 E\simeq 2\pi x_{\rm 4C}/L$ on the UV fixed point;\cite{Card84} we can
 numerically obtain the level by using discrete symmetries of the
 lattice Hamiltonian in the diagonalization calculations.
 Various excitations observed in TLL are characterized by a set of
 quantum numbers for symmetry operations.
 With respect to ${\cal O}_{\rm 4C}$, 
 it can be found in the subspace of
 the total spin $S^z_{\rm T}=0$ and
 the space inversion $P=-1$
 (the boundary condition is the same as that for the ground
 state).\cite{NakaEX}
 Suppose that $\Delta E(U,V,\delta,L)$ is a level corresponding to
 ${\cal O}_{\rm 4C}$ in the $L$-site system.
 Then, we numerically solve the phenomenological renormalization-group
 (PRG) equation $(L+2)\Delta E(U,V,\delta,L+2)=L\Delta E(U,V,\delta,L)$
 with respect to $\delta$ for given values of $U$ and $V$, where the gap
 behaves as $\Delta E\propto 1/L$ [i.e., an $L$-dependent transition
 point $\delta_\rho(U,V,L+1)$ (see Fig.\ \ref{FIG1})].\cite{Otsu02}
 After evaluating $\delta_\rho(U,V,L+1)$, we extrapolate them to the
 limit $L\to\infty$ using the formula
 $\delta_\rho(U,V,L)=\delta_\rho(U,V)+a L^{-3},$\cite{Itzy89} where
 $\delta_\rho(U,V)$ and $a$ are determined by the least-square-fitting
 condition.
\begin{figure}[t]
 \includegraphics[width=3.29in]{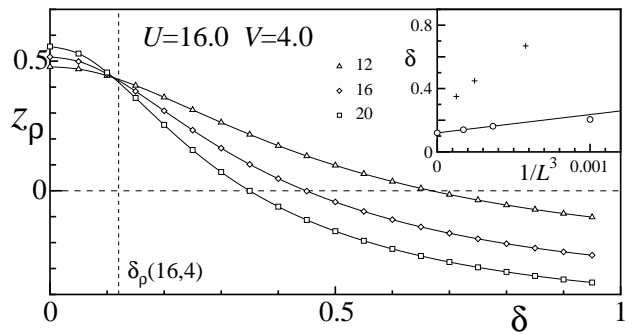}
 \caption{
 The $\delta$ dependence of $z_\rho(L)$ at $U=16$ and $V=4$.
 The vertical dotted line indicates the transition point (i.e., the PRG
 result) $\delta_\rho(16,4)\simeq 0.12$. 
 Inset plots $L$ dependences of $\delta'_\rho(U,V,L)$ (crosses) and
 the PRG data $\delta_\rho(U,V,L)$ (circles) with the fitting line.
 }
 \label{FIG2}
\end{figure}

 From the data of $L=12$-20, we obtain the phase boundary
 $\delta_\rho(U,V)$ as shown in Fig.\ \ref{FIG1}.
 We can check that, for all values of $U$ used here, the phase boundary
 lines converge to the point $(V,\delta)=(8,1)$ with the 2D-Ising
 criticality.
 On the other hand, while the finite-size corrections to the boundary
 may be large in weak-dimerization region, the boundaries also show
 convergences to the BKT-transition points $(V^*(U),0)$.
 Next we demonstrate the $\delta$ dependence of $z_\rho(L)$ in Fig.\
 \ref{FIG2}.
 With the increase of $\delta$, $z_\rho(L)$ decreases and becomes
 negative [we denote the zero point of $z_\rho(L)$ as $\delta'_\rho(U,V,L)$];
 this corresponds to the change of the center of mass as demonstrated in
 the above.\cite{Naka02}
 However, unlike, for instance, the Gaussian transition,
 $z_\rho(\infty)$ can take a finite value on the Ising transition point,
 so $\delta'_\rho(U,V,L)$ may not give an estimation of the transition
 point.
 In fact, the inset of Fig.\ \ref{FIG2} exhibits that $\delta'_\rho(U,V,L)$
 may be extrapolated to a value different from the PRG result.
 On the other hand, Fig.\ \ref{FIG2} also shows that there is a point
 $\delta\simeq 0.12$ at which $z_\rho(L)$ is almost independent of $L$.
 This crossing point is expected to be a good estimator for the Ising
 transition point because this is quite close to the PRG result even for
 small $L$.
 However, this issue remains as a future problem.

 Lastly, we shall refer to some implications of our study to the real
 materials.
 Besides the quantum-chemistry calculations,\cite{Duca86}
 the numerical estimations of the model parameters have been performed
 based upon the experimental data.\cite{Mila95,Nish00}
 For example, the realistic values of the dimerization parameter and the
 onsite Coulomb repulsion of (TMTTF)$_2$PF$_6$ have been estimated as
 $t_2/t_1\simeq 0.7$,
 $U/t_1\simeq 7.0$ ($t_{1,2}=1\pm \delta$),
 but the value of $V$ is still controversial
 (an uncertainty exists also in the value of
 $t_2/t_1$\cite{Nad_00,Frit91}).
 Our numerical estimation of the transition point using these values is
 $V_{\rm c}/t_1\simeq$ 4.0, 
 while generally the mean-field-type calculations tend to predict
 somewhat smaller values due to an overestimation of $V$
 effects.\cite{Seo_97,Shib01}
 On the other hand, several values have been reported for this material,
 e.g.,
 $V/t_1\simeq$ 2.8 (1.4) in
 Ref.\ \onlinecite{Mila95} (Ref.\ \onlinecite{Nish00}), 
 which is much smaller than the critical value, and thus predicts a
 uniform charge distribution
 (this conclusion may not be changed even in smaller dimerization
 cases).
 However, (TMTTF)$_2$PF$_6$ has the CO phase in the region above the
 lower-temperature spin-Peierls phase, and further it was theoretically
 suggested that a huge anomaly in the dielectric constant may reflect a
 nature of systems in the critical region.\cite{Tsuc02}
 This discrepancy may be attributed to many other interaction effects
 not included in the Hamiltonian.
 However, we think that since experimental findings seem to support the
 spin-charge separation with respect to the CO
 transition,\cite{Nad_00,Chow00} 
 the 1D electron models are to provide a primary description of real
 materials.


 To summarize, we investigated the ground-state phase diagram of the 1D
 quarter-filled extended Hubbard model with alternating transfer
 integral.
 Especially, the criticality on the phase boundary and the implication
 to the CO transition observed in the charge-transfer organic salts were
 mainly argued.

 One of the authors (H.O.) would like to thank 
 K. Mizoguchi
 and 
 Y. Okabe
 for stimulating discussions.
 M.N. is partly supported by the Ministry of Education, Culture, Sports,
 Science and Technology of Japan through Grants-in-Aid No.\ 14740241.
 Main computations were performed using the facilities of
 Yukawa Institute for Theoretical Physics, 
 and 
 the Supercomputer Center, Institute for Solid State Physics, University
 of Tokyo.

\vspace*{-5mm}

\end{document}